\documentclass[twocolumn]{jpsj2}

\title{Quantum spin chains with site dissipation}

\author{\textsc{Philipp Werner}$^{1}$, \textsc{Matthias Troyer}$^{1,2}$ and \textsc{Subir Sachdev}$^{3}$}
\inst{$^{1}$Institut f{\"u}r theoretische Physik, ETH H{\"o}nggerberg, CH-8093 Z{\"u}rich, Switzerland \\
$^{2}$Computational Laboratory, ETH Z\"urich, CH-8092 Z{\"u}rich, Switzerland \\
$^{3}$Department of Physics, Yale University, P.O. Box 208120, New Haven, CT 06520-8120, USA}
\date{\today}

\abst{ We use Monte Carlo simulations to study chains of Ising- and XY-spins with dissipation
coupling to the site variables. The
phase diagram and critical exponents of the dissipative Ising
chain in a transverse magnetic field have been computed
previously, and here we consider a universal ratio of
susceptibilities. We furthermore present the phase diagram and
exponents of the dissipative XY-chain, which exhibits a
second order phase transition. All our results compare well with
the predictions from a dissipative $\phi^4$ field theory.

}

\kword{spin chains, dissipation, phase transition, Monte Carlo}

\begin{document}

\maketitle

Dissipation and decoherence in mesoscopic quantum systems have
important practical implications, from stabilizing
superconductivity in granular materials, to the loss of
information stored in qubits. The macroscopic degree of freedom
can often be described by a spin variable of Ising or XY symmetry.
They might represent the lowest energy states in a double well
potential, or the phase of the superconducting order parameter on
some grain or island. Following Caldeira and Leggett
\cite{Caldeira} one can introduce dissipation in such a system by
coupling this spin variable linearly in the coordinates to an
infinite set of harmonic oscillators which represent the
environment. The spectral distribution of these harmonic
oscillators defines the type of dissipation and we will consider a
so-called Ohmic heat bath which produces linear friction. Upon
integrating out the heat bath degrees of freedom one obtains an
imaginary time effective action with long ranged interactions in
time.

The simplest of these models, the dissipative two-state (Ising)
system, has been studied extensively, both theoretically and by
numerical simulations \cite{Froehlich,Leggett,
Luijten&Messingfeld}. It exhibits a dynamical phase transition to
a localized state at a critical value of the dissipation strength.
We have recently studied what happens when infinitely many such
two-states systems are spatially coupled \cite{ising}. We observed
a quantum phase transition in a universality class different from
either the single site case or the non-dissipative system.

The second class of models we shall study involve `spin' degrees
of freedom with a global U(1) rotation (XY) symmetry. A single
XY-spin coupled to dissipation describes the physics of the
so-called single-electron box \cite{Schoen&Zaikin, ebox}. The
integer variable conjugate to the compact phase represented by the
spin corresponds to the number of excess charges on the small
metallic island, which is connected to an outside lead by a tunnel
junction. XY spins on higher dimensional lattices describe arrays
of Josephson junctions. In resistively shunted junctions, the
dissipation couples to the phase difference and the phases are
usually treated as non-compact variables. Such systems exhibit a
dissipation driven superconductor-to-metal transition
\cite{Schmid, Chakravarty&Ingold1, Korshunov1, junction} because the
dissipation suppresses phase slips and can restore
superconductivity. Very recently, an XY-chain with site
dissipation was studied in connection with the
superconductor-to-metal transition in nanowires\cite{Sachdev}.

In this work, we shall consider chains of Ising- and XY-spins with Ohmic dissipation coupling to the site variables. The discretized action for a chain of $n$-component spins with on-site dissipation reads
\begin{eqnarray}
S&=& -K_x\sum_{i=1}^{N_x}\sum_{\tau=1}^{N_\tau}\sigma_a(i,\tau)\sigma_a(i+1,\tau)\nonumber\\
&&-K_\tau\sum_{i=1}^{N_x}\sum_{\tau=1}^{N_\tau}\sigma_a(i,\tau)\sigma_a(i,\tau+1)\nonumber\\
&&-\alpha\sum_{i=1}^{N_x}\sum_{\tau<\tau'}
\Big(\frac{\pi}{N_\tau}\Big)^2\frac{\sigma_a(i,\tau)\sigma_a(i,\tau')}{(\sin(\frac{\pi}{N_\tau}|\tau-\tau'|))^2},
\label{chain_action}
\end{eqnarray}
with $a=1,\ldots,n$ and $\sum_a \sigma_a^2=1$ on every site. An Ising chain (in a transverse magnetic field) corresponds to $n$=1, a chain of XY-spins to $n=2$, and so on. The number of lattice sites in the spatial and imaginary time directions are $N_{x}$ and $N_{\tau}$, respectively, and we apply periodic boundary conditions in both directions.  The long-ranged term containing the dimensionless variable $\alpha$ describes the Ohmic dissipation coupling to the site variable $\sigma$.

The nature of the phase transition and its universal properties have recently been discussed by Pankov {\it et al.} \cite{Pankov}, who anayzed the dissipative $n$-component  $\phi^{4}$ field theory (the chain (\ref{chain_action}) corresponds to spatial dimension $d=1$)
\begin{eqnarray}
S &=& \int d^dx d\tau\left[\frac{1}{2}(\partial_x\phi_a)^2+\frac{s}{2}\phi_a^2+\frac{u}{24}(\phi_a^2)^2\right]\nonumber\\
&&+\frac{1}{2}\int d^dx d\omega|\omega||\phi_a(x,\omega)|^2.
\label{field_theory}
\end{eqnarray}
They predict a second order phase transition for $\alpha>0$, whose properties should not depend on the dissipation strength. The upper critical dimension of the field theory (\ref{field_theory}) is 2, independent of $n$, and an expansion to second order in $\epsilon=2-d$ yields the following expressions for the critical exponents \cite{Pankov, Sachdev}
\begin{eqnarray}
\nu &=& \frac{1}{2}+\frac{(n+2)}{4(n+8)}\epsilon\nonumber\\
&&+\frac{(n+2)(n^2+n(38-\frac{7\pi^2}{6})+132-\frac{19\pi^2}{3}))}{8(n+8)^3}\epsilon^2,\nonumber\\
\eta &=& \frac{(n+2)(12-\pi^2)}{4(n+8)^2}\epsilon^2, \nonumber\\
z &=& 2-\eta.
\label{epsilon}
\end{eqnarray}

An additional theoretical prediction concerns the Fourier
transform of the spin-spin correlation function, or susceptibility
$\chi(k,i\omega_n)$. Here, $k$ denotes a wave vector and
$\omega_n=2\pi n/N_\tau$ is the Matsubara frequency (the absolute
temperature is $T=\hbar/(k_B N_\tau)$). The strong hyperscaling
properties of the field theory in Eq.~(\ref{field_theory}) have
interesting consequences for its universal properties at the
quantum critical point. In particular, because the last $|\omega|$
term in Eq.~(\ref{field_theory}) does not renormalize, there is no non-universal cut-off
dependence to the overall scale of $\chi (k, \omega_n)$, which
thence obeys the scaling form
\begin{equation}
\chi(k, \omega) = \frac{\hbar}{k_B T} \Phi \left( \frac{c_1
k}{T^{1/z}}, \frac{\hbar \omega}{k_B T} \right),
\label{hyperscale}
\end{equation}
where only the number $c_1$ is non-universal, and all other
aspects of the scaling function $\Phi$ are independent of cut-off
scale parameters. In the present situation, we are not simulating
the field theory in Eq.~(\ref{field_theory}), but rather the
lattice Ising model in Eq.~(\ref{chain_action}), and so we need a
relation between the overall scale of $\phi$ and the Ising spins.
We sidestep the ignorance of the latter scale by computing the
ratio
\begin{equation}
\Upsilon(n)=\frac{\chi(0, i\omega_n)}{\chi(0, 0)}. \label{upsilon}
\end{equation}
In the limit of an infinite spatial length ($1 \ll N_\tau \ll
N_x^2$), Eq.~(\ref{hyperscale}) implies the remarkable result
that the function $\Upsilon (n)$ is a completely universal
function of the integer $n$. We can also determine the limiting
behavior of this function for small and large $n$. At low
frequencies, $\hbar \omega \ll k_B T$, we expect dissipative
behavior with a linear $|\omega_n|$ dependence of the dynamic
susceptibility. At large $\hbar \omega \gg k_B T$, scaling implies
\cite{Sachdev_book} $\chi (0, i\omega_n) \sim
1/|\omega_n|^{(2-\eta)/z} \sim 1/|\omega_n|$. Hence we have
\begin{equation}
\Upsilon(n) = \left\{\begin{array}{ll} A/n & \text{for } 1 \ll n \ll N_\tau, \\
1-n/B & \text{for } n \ll 1,
  \end{array}\right.
\label{upsilon_limit}
\end{equation}
where $A$ and $B$ are universal numbers.

We will first test the universality of the function $\Upsilon$ for
Ising spins and then present the phase diagram and critical
exponents for the XY-chain. In the Monte Carlo simulations, we use
a variant\cite{Luijten&Bloete} of the Swendsen-Wang algorithm
\cite{Swendsen&Wang} which builds the clusters in a time
$O(N_\tau\log N_\tau)$ despite long ranged interactions.

Since $n \ll 1$ is not accessible in imaginary time, we consider the case $1 \ll n \ll N_\tau$ and compute the amplitude $A$ for weak, intermediate and strong dissipation, setting $K_\tau=-\frac{1}{2}\ln[\tanh(1)]=0.136$ as in our recent study of the Ising chain\cite{ising}. For this value of $K_\tau$, the phase transition for a single dissipative Ising spin occurs at $\alpha_c=0.625$. The universal function $\Upsilon$, however, should not depend on the choice of $K_\tau$.

\begin{figure}[t]
\centering
\includegraphics[angle=-90, width=8.5cm]{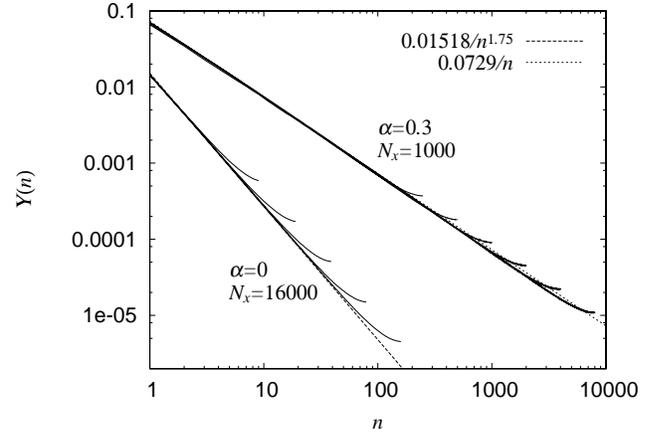}
\caption{$\Upsilon(n)$ for $\alpha=0$ and $\alpha=0.3$. The different curves correspond to different $N_\tau$ in the range $1 \ll N_\tau \ll N_x$ ($\alpha=0$) and $1 \ll N_\tau \ll N_x^2$ ($\alpha=0.3$) respectively. The data are plotted in the range $1 \le n \le N_\tau/2$.}
\label{fig:full_range}
\end{figure}
\begin{figure}[t]
\centering
\includegraphics[angle=-90, width=8.5cm]{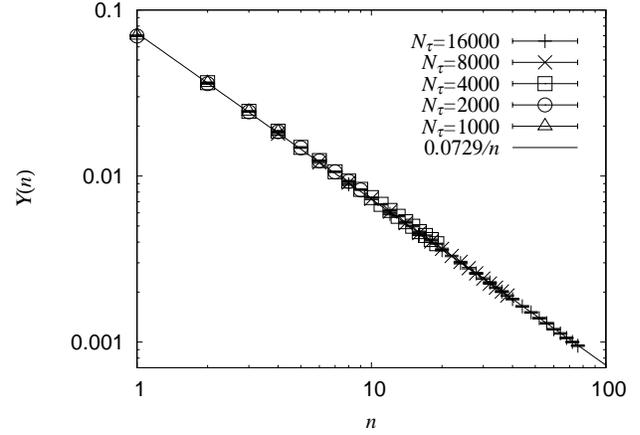}
\caption{$\Upsilon(n)$ for $\alpha=0.3$ and $\frac{1}{1000}\frac{N_\tau}{2} \le n \le \frac{1}{100}\frac{N_\tau}{2}$. The data sets corresponding to different $N_\tau$ are plotted in different colors for better clarity. The dotted line shows the best fit to $A/n$.}
\label{fig:restricted_range}
\end{figure}

For $\alpha=0$ (non-dissipative 2D Ising model), the amplitude $A$
can be calculated exactly. One finds for large $n$
\cite{Sachdev_book}
\begin{equation}
\Upsilon(n)\sim\left(\frac{2}{n}\right)^{7/4}\left(\frac{\Gamma(15/16)}{\Gamma(1/16)}\right)^2=\frac{0.01518}{n^{1.75}},
\end{equation}
which we used to test the numerical simulations. (Note that the
large $n$ behavior $\sim n^{-7/4}$ applies only for
$\alpha=0$; any non-zero $\alpha$ is expected to crossover into
the behavior in Eq.~(\ref{upsilon_limit}).) In this case the
dynamical critical exponent is $z=1$, so we plot in
Fig.~\ref{fig:full_range} the function $\Upsilon(n)$, computed
at the critical inter-site coupling $K_x^c$, for different
lattices with $1 \ll N_\tau \ll N_x$.  The behavior predicted from
the Onsager solution is nicely reproduced in the scaling limit $1
\ll n \ll N_\tau$. Also shown in Fig.~\ref{fig:full_range} are
the data for $\alpha=0.3$. Since $z\approx 2$ for $\alpha>0$, we
chose lattices with $1\ll N_\tau \ll N_x^2$. A power-law decay
proportional to $1/n$ is clearly visible, although the curves bend
away slightly for $n$ outsite the scaling region $1\ll n \ll
N_\tau$. For the purpose of extracting the amplitude $A$, we
define the scaling region as $\frac{1}{1000}\frac{N_\tau}{2} \le n
\le \frac{1}{100}\frac{N_\tau}{2}$.
Fig.~\ref{fig:restricted_range} shows the data for $\alpha=0.3$
restricted to these values of $n$. A fit to the predicted
power-law (\ref{upsilon_limit}) yields $A=0.0729$. Repeating the
same analysis for $\alpha=0.1$, we get $A=0.0726$. For very strong
dissipation ($\alpha=0.5$) we see stronger corrections to scaling
and $n\approx \frac{1}{100}\frac{N_\tau}{100}$ lies clearly
outside the scaling region. Larger lattices would be needed to
extract an accurate value of $A$. Nevertheless, even our data at
$\alpha=0.5$ are roughly consistent with the asymptotic form
\begin{equation}
\Upsilon(n) \sim \frac{0.073}{n}.
\end{equation}

Having confirmed the predictions for $\Upsilon$, we use quantum Monte Carlo simulations to determine the phase diagram and critical exponents of the dissipative XY-spin chain following the analogous investigation of the Ising-spin case in Ref. \citen{ising}. Assuming a second order phase transition, we use a scaling procedure which was proposed for the study of spin glasses \cite{Rieger}.
\begin{figure}[t]
\centering
\includegraphics [angle=-90, width= 8.5cm] {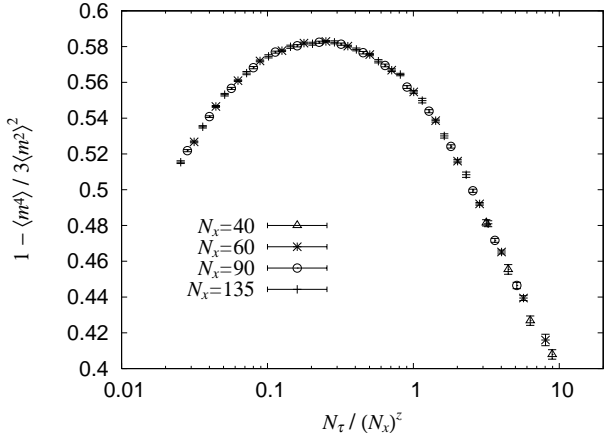}
\caption{Data collapse of Binder cumulants at the critical point for $\alpha=0.3$ and $z=1.97$.}
\label{fig:collapse}
\end{figure}
For a given value of $\alpha$ ($K_\tau=0.1$ is kept fixed in this study) we determine the critical spatial coupling $K_x^c$ and the dynamical critical exponent $z$ self-consistently from the data collapse of the Binder cumulant ratio $B=1-{\langle m^4\rangle}/({3\langle m^2\rangle ^2})$, where $m$ is the magnetization. Near the critical point, $B$ scales as
\begin{equation}
B(N_x, N_\tau) = \phi_B\Big(\frac{N_{x}}{\xi}, \frac{N_\tau}{(N_x)^z}\Big). \label{3.1}
\end{equation}
At the critical point, the correlation length $\xi$ diverges, and the Binder cumulants collapse onto a universal function of $N_\tau/(N_x)^z$. The excellent data collapse shown in Fig.~\ref{fig:collapse} validates the scaling {\em Ansatz} in Eq. (\ref{3.1}) and confirms the second order transition which was predicted in Pankov \textit{et al.}\cite{Pankov}. The data collapse yields $K_x^c=0.92132(2)$ and $z=1.97(3)$.

\begin{figure}[t]
\centering
\includegraphics [angle=-90, width= 8.5cm] {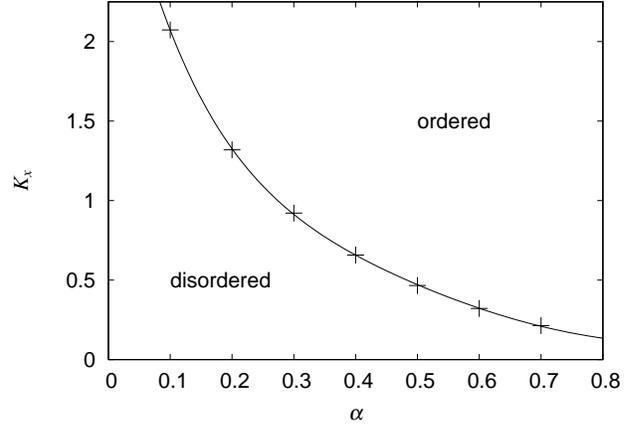}
\caption{Phase diagram of the dissipative XY-spin chain in the space of nearest neighbor coupling $K_x$ and dissipation strength $\alpha$ ($K_\tau=0.1$).}
\label{fig:phasediagram}
\end{figure}

Repeating this analysis for several values of $\alpha$, we map out the phase diagram shown in Fig.~\ref{fig:phasediagram}. In contrast to the Ising case \cite{ising, Luijten&Messingfeld, Froehlich} there exists no ordered state at $K_x=0$. As mentioned in the introduction, the single site case corresponds to the imaginary time effective action of a single-electron box \cite{Schoen&Zaikin}, which was recently studied using extensive Monte Carlo simulations \cite{ebox}.

\begin{figure}[htb]
\centering
\includegraphics [angle=-90, width=8.5cm] {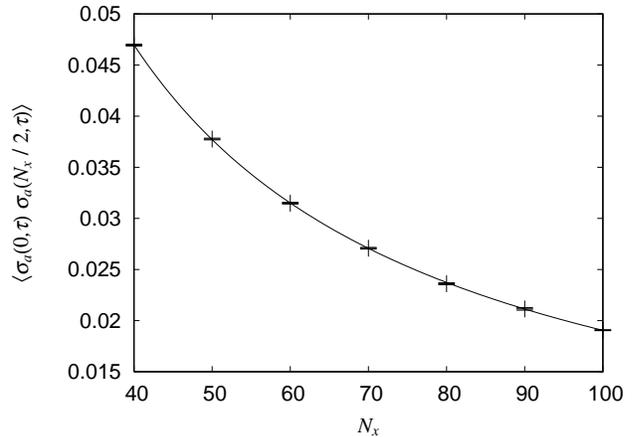}
\caption{Equal-time correlation $\langle \sigma_a(0,\tau)\sigma_a(N_x/2,\tau)\rangle$ at the critical point for $\alpha=0.3$. The time direction of the lattice is scaled as $N_\tau = 10 N_x^2$. The curve shows a power-law fit to the data.}
\label{fig:corr_L}
\end{figure}

Next, we compute the critical exponents $\eta$ and $\nu$. The order parameter correlation function $C(x,\tau)$ in an infinite system and at the critical coupling $K_x^c$ scales as\begin{eqnarray}
C(x,\tau) \sim x^{-(z+\eta -1)}\tilde{g}(\tau/x^z)=\tau^{-(z+\eta -1)/z}\hat{g}(x^z/\tau).
\label{4}
\end{eqnarray}
The equal-time correlation function at the critical point should therefore decay asymptotically as $C(x,0) \sim x^{-(z+\eta-1)}$. The corrections to scaling are larger than in the Ising case \cite{ising} and we cannot directly fit a periodic version of this power-law to the correlation function. Instead we consider the correlations between the most distant sites for different system sizes. In Fig.~\ref{fig:corr_L} we plot $\langle \sigma_a(0,\tau)\sigma_a(N_x/2,\tau)\rangle$ as a function of $N_x$. A power-law fit to these data yields $z+\eta=1.985(20)$.

\begin{figure}[t]
\centering
\includegraphics [angle=-90, width= 8.5 cm] {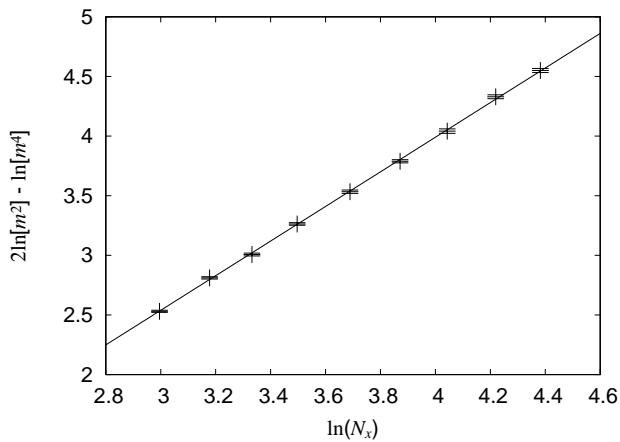}
\caption{Plot of $2\ln [m^2]-\ln [m^4]$ ($[m^n]$ is defined in (\ref{m_n})) as a function of $\ln N_x$ at the critical point ($\alpha=0.3$, $K_x=0.1$). We scaled the imaginary time dimension as $N_\tau=5(N_x)^z$ with z=2 and z=1.97 (not shown). The slope of the fitted line gives $1/\nu$.}
\label{fig:nu}
\end{figure}

In order to determine the critical exponent $\nu$, we set $N_\tau=5(N_x)^z$ to reduce the two-parameter scaling to a one-parameter one. This allows us to employ the procedure outlined in Refs.~\citen{Ferrenberg&Landau} and \citen{ising}. Using the definition
\begin{equation}
[m^n]=\langle m^n\Sigma_x\rangle_{K_x^c}-\langle m^n\rangle_{K_x^c}\langle \Sigma_x\rangle_{K_x^c},
\label{m_n}
\end{equation}
where $m$ denotes the magnetization and $\Sigma_x$ the spatial coupling energy, one finds from finite size scaling considerations
\begin{equation}
2\ln[m^2] - \ln[m^4] = \text{const}+\frac{1}{\nu}\ln N_x.
\label{nu}
\end{equation}
The left hand side of Eq.~(\ref{nu}) is plotted in Fig.~\ref{fig:nu} for $\alpha=0.3$ and $K_\tau=0.1$. From the slope of a linear fit to these data one obtains $\nu$. The major source of error is the uncertainty on $z=1.97(3)$. Hence, we determinded the slopes for $z=1.97$ and $z=2$ and find $\nu=0.689(6)$.

Having obtained the critical exponents for $\alpha=0.3$, we would like to compare our result with the predicitions from the dissipative $n$-component $\phi^{4}$ field theory (\ref{field_theory}). One dimensional chains correspond to $\epsilon=1$. In Table~\ref{tab:compare} we list the numerically determined exponents for Ising ($n=1$) and XY-chains ($n=2$) and the values from the two-loop $\epsilon$-expansion (\ref{epsilon}). The agreement is remarkably good. In the Ising case we found that the exponents do not depend on the dissipation strength and we expect the same for the XY-chain as well.

\begin{table}
\centering
\caption{Comparison of exponents obtained by our simulations and the values predicted by the $\epsilon$-expansion in order $O(\epsilon^2)$. The values for the Ising chain are from Ref.~\citen{ising}.}
\begin{tabular}{lllll}
  \hline
  \hline
  & Ising \cite{ising} & & XY & \\
  & simulation\hspace{0mm} & $\epsilon$-expansion\hspace{0mm} & simulation\hspace{0mm} &  $\epsilon$-expansion \\
  \hline
  $\nu$ & $0.638(3)$  & 0.633 & $0.689(6)$  & 0.663 \\
  $\eta$ & $0.015(20)$ & 0.020 & $0.015(45)$ & 0.021 \\
  $z$ & $1.985(15)$ & 1.980 & $1.97(3)$ & 1.979 \\
  $z+\eta\hspace{2mm}$ & $2.00(1)\hspace{4mm}$ & 2 & $1.985(20)\hspace{2mm}$ & 2 \\
  \hline
  \hline
\end{tabular}
\label{tab:compare}
\end{table}

M. T. and P. W. acknowledge support by the Swiss National Science Foundation. The calculations have been performed on the Asgard Beowulf cluster at ETH Z\"urich, using the open source ALPS library \cite{ALPS}.

\end{document}